\documentclass[12pt]{article}
\usepackage{graphicx}
\begin{document}
\rightline{NKU-2015-SF1}
\bigskip

\newcommand{\be}{\begin{equation}}
\newcommand{\ee}{\end{equation}}
\newcommand{\noi}{\noindent}
\newcommand{\refb}[1]{(\ref{#1})}
\newcommand{\ra}{\rightarrow}
\newcommand{\bib}{\bibitem}
\newcommand{\bigt}{\bigtriangleup}

\begin{center}
{\Large\bf Quasi normal modes and the area spectrum  of a near extremal de Sitter black hole with conformally coupled scalar field}

\end{center}
\hspace{0.4cm}
\begin{center}
Sharmanthie Fernando \footnote{fernando@nku.edu}\\
{\small\it Department of Physics \& Geology}\\
{\small\it Northern Kentucky University}\\
{\small\it Highland Heights}\\
{\small\it Kentucky 41099}\\
{\small\it U.S.A.}\\

\end{center}

\begin{center}
{\bf Abstract}
\end{center}

\hspace{0.7cm} 

In this paper we have studied a black hole in de Sitter space which has a conformally coupled scalar field in the background. This  black hole is also known as the $MTZ$ black hole. We have obtained exact values for the quasi normal mode frequencies under massless scalar field perturbations. We have demonstrated that when the black hole is near-extremal, that the wave equation for the massless scalar field simplifies to a Schr$\ddot{o}$dinger type equation with the well known P$\ddot{o}$shler-Teller potential. We have also used 6th order WKB approximation to compute quasinormal mode frequencies to compare with exact values obtained via the P$\ddot{o}$shler-Tell method for comparison.  As an application, we have obtained the area spectrum using modified Hods approach and show that it is equally spaced.

{\it Key words}: static, conformally, extreme, black hole, quasi-normal modes, area spectrum


\section{ Introduction}

No hair conjecture which says that black holes can be characterized only by its mass, charge and angular momentum can be violated by the presence of scalar fields. There are several examples of such hairy black holes with minimally coupled scalar field \cite{hen1} \cite{hen2} \cite{zan1} \cite{mart2}. There are also interesting black hole solutions with non-minimally coupled scalar fields.  A charged black hole with  scalar hair where  the scalar field was nonminimally coupled to gravity was presented by Xu amd Zhao \cite{xu1}.

In this paper, we are interested in a black hole solution with conformally coupled scalar field called MTZ black  hole constructed by Martinez et.al \cite{mart1}. This particular black hole has a self interacting scalar field which is real and is non-minimally coupled to gravity. MTZ black hole is asymptotically de Sitter and has a scalar field which diverge inside the horizon. Thermodynamics of the MTZ black hole was studied by Barlow et.al. \cite{eli}.  Instability of the MTZ black hole was studied by Harper et.al \cite{harper}.

There are several interesting works on black holes with a conformal scalar field. Anabal$\acute{o}$n and Maeda \cite{ana1} obtained a class of black holes which belong to the Pleba$\acute{n}$ski-Demia$\acute{n}$ski family which contain the MTZ black hole as a subclass.  The issue of existence of exact black hole solutions and their thermodynamical properties for spacetimes with a conformally coupled scalar field in $n$ dimensions were discussed by Nadalini et.al \cite{nada1}.  An asymptotically anti de Sitter black hole in 2+1 dimensions with a conformally coupled scalar field was found in \cite{mart3}.

In this paper, we are interested in studying the quasinormal modes (QNM) of a near extreme MTZ black hole and deriving the area spectrum from the frequencies of the QNM. When a black hole is perturbed, after the initial response, the black hole starts to vibrate with frequencies which are complex. Such oscillations are called quasinromal modes (QNM). QNM frequencies are insensitive to the initial perturbation and only depend on the properties of the black hole such as mass, charge and angular momentum. QNM have attracted lot of attention in black hole physics research community. Two main reasons are the relation of QNM's to gravitational wave astrononomy \cite{ferra} and the AdS/CFT correspondence \cite{alex}. We will omit getting into details of applications of QNM and refer to excellent reviews by Berti et.al. \cite{cardoso1} and Konoplya and Zhidenko \cite{kono1}.

QNM also have attracted attention due to its relation to the area spectrum of a quantum black hole. The history of this aspect starts with Bekenstein's famous conjecture \cite{bek1} \cite{bek2} that in a quantum theory, the black hole horizon area should have a equally spaced discrete spectrum such that,
\be
A_n = \epsilon l_P^2 n; \hspace{1 cm} n=0,1,2,...
\ee
Here, $l_P$ is the Plank length and we have taken $ c = G = 1$. Therefore, $l_P^2 = \hbar$. The exact value of $\epsilon$ has to be determined by a quantum theory of gravity. Hod made an interesting argument that $\epsilon$ could be determined by the real part of the QNM's for high overtones \cite{hod}. This is known as  Hods conjecture. Specifically, it relates the fundamental mass unit in a quantum theory to the real part of QNM frequency ($\omega_R$) as,
\be \label{hod2}
\bigt M = \hbar \omega_R
\ee
By applying this conjecture to the Schwarzschild black hole, Hod determined $\epsilon = 4 ln 3$.

Maggiore \cite{magi} gave a new interpretation to QNM by arguing that a perturbed black hole should be treated as a collection of harmonic oscillators. Also he argued that  the frequency of the oscillations should be,
\be
 \omega_p = \sqrt{ |\omega_R|^2 + |\omega_I|^2 }
 \ee
Since  for high overtone QNM (large n),   $|\omega_I| \gg \omega_R$,  in the large $n$ limit, $\omega_p \approx |\omega_I|$ rather than $\omega_R$.  Also $\omega_R$ in eq.$\refb{hod2}$ should be replaced by  the transition between adjacent QNM frequencies such that,
\begin{equation} \label{omegac2}
\bigtriangleup \omega  = \lim_{n \rightarrow \infty}  \omega_p( n + 1 ) - \omega_p(n)
\end{equation}
Hence, the modified Hod's conjecture can be represented as
\be \label{hod}
\bigtriangleup M  =  \hbar \left(\lim_{n \rightarrow \infty}  \omega_p( n + 1 ) - \omega_p(n)\right) \approx \hbar lim_{n \ra \infty} \left(|(\omega_I)_n| - |(\omega_I)_{n -1}| \right)
\ee
Using this modified approach, $\epsilon$ was found to be $8 \pi$ and the area spectrum was found to be equally spaced for the Schwarzschild black hole \cite{magi}. Vagenas showed that the area spectrum if equally spaced and $A_n = 8 \pi \hbar n$ for the Kerr black hole using the modified Hods conjecture \cite{elias}. L$\acute{o}$pez-Ortega demonstrated that a $d$ dimensional de Sitter space have an area spectrum $A_n = 16 \pi \hbar n$ with $\epsilon = 16 \pi$ \cite{lopez}.

In this paper, we will explore the area spectrum of the near extreme MTZ black hole via the modified Hod's approach.

The paper is structured as follows: In section 2 we will introduce the MTZ  black hole. In section 3 we will discuss the properties of the near extreme black hole. In section 4, we will discuss the massless scalar field perturbation of the black hole and derive exact values for the QNM frequencies for the near extreme MTZ black hole. In section 5, we will discuss the area spectrum and finally in section 6 the conclusion is given.


\section{ Introduction to the MTZ black hole}

In this section, we will give an introduction to the MTZ black hole. The action for  a scalar which is conformally coupled to gravity with  a positive cosmological constant is given by,
\be
S = \int d^4x \sqrt{ -g} \left( \frac{ R - 2 \Lambda}{ 16 \pi}  -\frac{1}{2} \partial^{\mu} \Phi \partial_{\mu} \Phi  - \frac{ R \Phi^2}{12} - \alpha \Phi^4 \right)
\ee
We have taken $ G = c = 1$. $R$ is the scalar curvature, $\Lambda$ is the cosmological constant, $\Phi$ is the scalar field and $\alpha$ is a coupling constant.  In the above action, the matter part is invariant under the conformal transformation,
\be
g_{\mu \nu} \ra \Omega(x)^2 g_{\mu \nu}; \hspace{1 cm} \Phi \ra  \frac{ \Phi}{ \Omega(x)}
\ee
MTZ black hole derived by Martinez et.al. \cite{mart1} is given by
\be \label{metric}
ds^2 = -f(r) dt^2 + \frac{dr^2}{f(r)}  + r^2 ( d \theta^2 + sin^2 \theta d \phi^2)
\ee
where
\begin{equation}
f(r) = \left(1 - \frac{ M} { r} \right)^2   - \frac{ r^{ 2}}{ l^2}
\end{equation}
The scalar field $\Phi$ is given by,
\be \label{scalar}
\Phi = \sqrt{ \frac{3}{ 4 \pi} } \left( \frac{ M}{ r - M} \right)
\ee
The scalar curvature $R = 4 \Lambda$. Note that $\Lambda = \frac{3}{l^2}$. The given solution exists only  if,
\be
\alpha = -\frac{2}{ 3 l^2}
\ee
One may notice that the geometry given by the metric in eq.$\refb{metric}$,  is the same as the Reissner-Nordstrom-de Sitter black hole with $M =Q$.

The MTZ black hole have three horizons, inner, event and cosmological horizons respectively given by,
\be
r_i = \frac{l}{2} \left[ -1 + \sqrt{ 1 + \frac{ 4 M}{l}} \right]
\ee
\be
r_e = \frac{l}{2} \left[ 1 - \sqrt{ 1 - \frac{ 4 M}{l}} \right]
\ee
\be
r_c = \frac{l}{2} \left[ 1 + \sqrt{ 1 - \frac{ 4 M}{l} }\right]
\ee
There is a curvature singularity at $ r =0$. The scalar field is singular at $ r = M$. It is clear that when $ M = \frac{l}{4}$, $ r_e = r_c$. Such space-times are called Nariai black holes \cite{fernando1}. Hence, the event and cosmological horizon exists only if $ M < \frac{l}{4}$.

The Hawking temperature of the MTZ black hole is,
\be
T_H = \frac{1}{ 2 \pi l} \sqrt{ 1 - \frac{ 4 M}{ l} }
\ee
Hawking temperature is the same for both the event and cosmological horizon. Hence, from the point of view of the Reissner-Nordstrom-de Sitter black hole, such space-times are called lukewarm black holes \cite{romans}.

The entropy of the black hole does not obey the usual area law. Instead, it is given by \cite{eli}
\be \label{entropy}
S_{ e,c} = \pi r_{e,c}^2 \left( 1 - \frac{1}{6} \Phi(r_e, r_c)^2 \right)
\ee


\section{ Near-extremal black holes}

In this paper, we focus on the near extremal MTZ black hole. In the   near extremal black hole  the black hole event horizon $r_e$ is very close to the cosmological horizon $r_c$. Lets define the extreme black hole radius as $r_{ex} = r_c = r_e = \frac{l}{ 2}$. Note that this occurs when $ M = \frac{l}{4}$. Now, when $ r_e \approx r_c$, one can expand the function $f(r)$ in a Taylor series as \cite{paw}
\begin{equation}
f(r) = \frac{ f''(r_{ex})}{2} ( r - r_e) ( r - r_c)
\end{equation}
Here,
\be
f''(r) = \frac{ d^2 f(r)}{ dr^2}
\ee
For the MTZ black hole,
\be
f''(r_{ex}) = -  \frac{4}{l^2} 
\ee
For  the near extreme MTZ black hole the tortoise coordinate $r_*$ can be defined as,
\be \label{tor2}
r_* = \int \frac{ dr}{ f(r)} =   \frac{l^2}{ 2 ( r_e - r_c)}   log\left( \frac{ r_c - r}{ r - r_e} \right) 
\ee
When $r \ra r_e$, $r_* \ra -\infty$ and $ r \ra r_c$, $r_* \ra \infty$. 
\noindent
One can invert the relation in eq.$\refb{tor2}$ to obtain $r$ as,
\be \label{rvalue}
r = \frac{ r_e + r_c e^{ \eta r_*}}{ 1 + e^{ \eta r_*}}
\ee
Here,
\be \label{eta}
\eta = \frac{2}{l^2}  ( r_c - r_e) = \frac{2}{l^2} \sqrt{ 1 - \frac{4 M}{l}}
\ee
With the  definition given in eq.$\refb{rvalue}$ for $r$, the function $f(r)$ for the near extream MTZ black hole can be written as,
\be \label{newfr}
f(r)  =  \frac{ \gamma}{ 4( Cosh( \frac{ \eta r_*}{2}))^2} 
\ee
Here,
\be
\gamma = \frac{2}{l^2} ( r_c - r_e)^2= 2 ( 1 - \frac{4 M}{l})
\ee


\section{ Quasi normal modes of a massless scalar perturbation of a black hole}

In this paper, we are interested as to how the MTZ black hole  reacts to perturbations by a massless scalar field. First  we will give an introduction to the scalar field perturbations by  a  black hole. The equation for a massless scalar field around a  black hole is given by the Klein Gordon equation,
\be \label{klein}
\bigtriangledown^2 \Psi =  0
\ee
With the ansatz for the scalar field given by,
\be
\Psi = e^{ - i \omega t}  Y_{ j, m} ( \theta, \phi) \frac{ R(r)}{ r}
\ee
eq.$\refb{klein}$ simplifies to a Schr$\ddot{o}$dinger-type equation given by,
\be \label{wave}
\frac{ d^2 R(r_*) }{ dr_*^2} + \left( \omega^2 - U_{eff}(r_*)  \right) R(r_*) = 0
\ee
with,
\be \label{pot}
U_{eff}(r_*) = \frac{ j ( j + 1) f(r)} { r^2} + \frac{ 1}{ 2 r}  \frac{ d( f(r)^2)}{dr}
\ee
Here, $j$ is the spherical harmonic index and  $r_*$ is the tortoise coordinates  defined in section(3) in eq.$\refb{tor2}$.

Now we can write the effective potential  in eq.$\refb{pot}$  for the near extreme MTZ black hole with $f(r)$ given in eq.$\refb{newfr}$. We have assumed $f'(r) \approx 0$ since the black hole is near extremal. Therefore, only the first term is dominant in the effective potential. We have also approximated $r \approx r_{e}$. With these approximations, the effective potential can be written as,
\be
U_{eff} =  \frac{ U_0}{ Cosh^2( \frac{\eta r_*}{2})}
\ee
Here,
\be \label{v0}
U_0 = \frac{ \gamma j ( j+1)}{ 4 r_{e}^2} = \frac{2 ( 1 - \frac{4 M}{l}) j ( j +1)}{ 4 r_e^2}
\ee
Now the wave equation for the scalar perturbations simplifies to,
\be \label{wave2}
\frac{ d^2 R(r_*) }{ dr_*^2} + \left( \omega^2 -   \frac{ U_0}{ Cosh^2( \frac{\eta r_*}{2})}  \right) R(r_*) = 0
\ee
The potential $\frac{ U_0}{ Cosh^2( \frac{\sigma r_*}{2})}$ is the well known   P$\ddot{o}$shl-Teller potential. Ferrari and Mashhoon \cite{mash} demonstrated that  $\omega$ can be computed exactly for the  wave  eq.$\refb{wave2}$ as,
\be
\omega =  \sqrt{ U_0  - \frac{\eta^2}{16}} - i \frac{\eta}{2} ( n + \frac{1}{2})
\ee
Here $U_0$ and $\eta$ are given in eq.$\refb{v0}$ and eq.$\refb{eta}$.

We have made comparisons with  $\omega$ values obtained via  the 6th order WKB method developed by Konoplya \cite{kon}  and the ones  obtained using the P$\ddot{o}$shl-Teller approximation. The results are displayed in Table 1. One can observe that both approaches agree. $\omega_I$  agree better compared with  $\omega_R$. The  agreement give validation for the approximation we have used for the near extreme black hole.

\begin{center}
\begin{tabular}{|l|l|l|l|l|r} \hline \hline

 j & n & $l$  & $\omega$(P-T)  &  $\omega$ (WKB)  \\ \hline

1 & 0 & 4.1 &  0.0883 - i 0.0190  &  0.0755 - i 0.0196 \\ \hline

2 & 0 & 4.1 & 0.1552 - i 0.0191 & 0.1332 - i 0.0195 \\ \hline

2 & 1 & 4.1 &  0.1552 - i 0.0571 & 0.1330 - i 0.0584 \\ \hline

3 & 1 & 4.1 & 0.2203 - i 0.0571 & 0.1891 - i 0.0583 \\ \hline

3 & 2 & 4.1 & 0.2203 - i 0.0952 & 0.1888 - i 0.0972 \\ \hline

4 & 2 & 4.1 & 0.2849 - i 0.0952 & 0.2444 - i 0.0971 \\ \hline
\end{tabular}
\end{center}

Table 1. The table shows $\omega$ from the P$\ddot{o}$shl-Teller approximation and WKB method. Here, $ M =  1$. Note that for extreme black hole, $  l = 4 M = 4$. Hence, we have chosen a $l$ value close to $4$ for comparison.\\

\section{ Area spectrum of the near-extreme MTZ black hole}

In this section we will apply modified Hods conjecture described in the introduction to derive the area spectrum of the near-extreme MTZ black hole. Let us recall the modified Hods conjecture as,
\be
\bigtriangleup M = \hbar lim_{n \ra \infty}  \left(|(\omega_I)_n| - |(\omega_I)_{n -1}| \right) 
\ee
Since $\bigt M = \bigt A \left( \frac{ dM}{ d A} \right)$
\be \label{area}
\bigtriangleup A = \hbar  lim_{n \ra \infty} \left(|(\omega_I)_n| - |(\omega_I)_{n -1}| \right) \left( \frac{ dA}{ d M} \right)
\ee
Since $A = 4 \pi r_e^2$, for the MTZ black hole,
\be 
\left( \frac{ dA}{ d M} \right) =  8 \pi r_e^2 \frac{ 1} { \sqrt{1 - \frac{ 4 M}{ l} }}
\ee
For the MTZ black hole,
\be
\left(|(\omega_I)_n| - |(\omega_I)_{n -1}| \right) = \frac{ \eta}{2}
\ee

By substituting  the above facts to the  eq.$\refb{area}$ 
\be
\bigtriangleup A = 8 \hbar \pi \left( 1 - \sqrt{ 1 - \frac{4M}{l}}\right)
\ee
Since the black hole is near extreme, $ M \approx \frac{l}{ 4}$. Therefore,
\be
\bigtriangleup A = 8 \hbar \pi 
\ee
Since $ \bigtriangleup A = A_{n+1} - A_n$, 
\be
A_n = 8 \hbar \pi n
\ee
From the results obtained above, it is obvious that the area spectrum is equally spaced for the near-extreme MTZ black hole.  
Since $A_n = 8 \pi \hbar n = 4 \pi r_e^2$, $r_e$ also gets quantized as,
\be
r_{e(n)} = \sqrt{ 2 \hbar n}
\ee
$r_e$ has a discrete spectrum but is not equally spaced. Due to the fact that $M$ and $r_e$ are related, the mass $M$ can be shown to be quantized as,
\be
M_n = \sqrt{ 2 \hbar n} - \frac{ 2 \hbar n}{ l}
\ee
The scalar field on the horizon can be computed via eq.$\refb{scalar}$,
\be
\Phi_n = \frac{ \sqrt{3}}{ \sqrt{ 4 \pi}} \left( \frac{ l}{ \sqrt{ 2 \hbar n}} -1 \right)
\ee
The entropy can be calculated from eq.$\refb{entropy}$,
\be
S_n = \frac{A_n}{4} \left( 1 - \frac{\Phi(r_e)_n^2}{6} \right) = 2 \pi \hbar n - \frac{ \hbar n}{4} - \frac{l^2}{8} + \frac{ l \sqrt{ \hbar n}}{ 2 \sqrt{2}}
\ee
it is clear that the entropy does not equally spaced. There are extra terms other than the usual $ 2 \pi \hbar n$ term one expects.


\section{ Conclusion}
In this paper, we have studied QNM  of  near-extreme MTZ black hole. MTZ black hole is obtained with a conformally coupled scalar field in its background and it is asymptotically de Sitter. The MTZ black hole has the same geometry as the lukewarm Reissner-Nordstrom-de Sitter black hole.  The MTZ black hole has two  horizons; an event horizon $(r_e)$ and a cosmological horizon $(r_c)$. For the near-extreme black hole, $ r_e \approx r_c$. The massless scalar field perturbation for the near extreme MTZ black hole is shown to simplify to a Schr$\ddot{o}$dinger-type wave equation with a  P$\ddot{o}$shl-Teller potential. Since the Schr$\ddot{o}$dinger  equation with a P$\ddot{o}$shl-Teller potential can be solved exactly, we found the QNM frequencies exactly for the near extreme MTZ black hole. To validate our approximation, we used  6th order $WKB$ approach developed by Konoplya \cite{kon} to compute $\omega$ .  There is  a remarkable agreement between the two methods: $\omega_I$ values are much more closer than $\omega_R$.

We applied the modified Hod's conjecture to obtain the area spectrum. We found  that the area spectrum is  equally-spaced and given by $A_n = 8 \pi \hbar n$. The mass and the entropy of the black hole also are discrete but they are not equally spaced.

The conclusion will not be complete if we do not make remarks of the well known Kunstatter's method used to find the area spectrum in the literature \cite{kun}. Kunsttater proposed an  adiabatic invariant quantity $I$  given by,
\be \label{kun}
I = \int  \frac{ dE}{ \omega}
\ee
Here, $E$ is the energy of the black hole and $\omega$ is the QNM frequency. For  large $n$, due to the Bohr-Sommerfeld quantization, $I$ approximate to,
\be
I \approx n \hbar, \hspace{1 cm} n\ra \infty
\ee
The method of Kunstatter, with $\omega$ in eq.$\refb{kun}$ as  $\bigtriangleup \omega$ in eq.$\refb{omegac2}$  has been employed to obtain the area spectrum of many black holes \cite{elias}\cite{wei2}\cite{fernando6} \cite{ranli}\cite{kwon} \cite{bertha}\cite{yun}. However, Kunstatter's approach does not work well  for near extreme black holes as  described   in  \cite{wei2} \cite{med}. That is  the reason  only the Modified Hod's approach to obtain the area spectrum for the near extreme black hole considered in this paper.

In light of the results we obtained in this paper for the MTZ black hole, it is important to compare the results obtained for other well known black holes along those lines.  First, we like to make comparisons with the well known BTZ black hole in 2+1 dimensions: Setare \cite{seta} employed Kunstatter's method to compute the area spectrum of the non-rotating BTZ black hole. He used $\omega_R$ in the expression $I = \int  \frac{ dE}{ \omega} $ and the resulting area spectrum was shown to be non-equally spaced.  Kwon and Nam \cite{nam1}, used Maggiore's approach and used  $\bigt \omega \approx |\omega_I|_{n+1} - |\omega_I|_n$ in $ I = \int \frac{ dM}{ \bigt \omega}$  and  obtained equally spaced area spectrum for the non-rotating BTZ black hole. We went one step further and used the modified Hod's approach to compute the area spectrum of the non-rotating BTZ black hole and saw that the area spectrum is indeed equally spaced. The area spectrum of the extreme Reissner-Nordstrom black hole was computed by Setare using $\omega_R$ in the expression $I = \int  \frac{ dE}{ \omega} $ and demonstrated that it has equally spaced area spectrum \cite{seta3}. Setare and Vagenas investigated the area spectrum of the Kerr and the extremal Kerr black holes in \cite{seta4}: the Kerr black hole area spectrum was not equally spaced while the extremal Kerr black hole did. Area spectrum of the Schwarzschild-de Sitter black hole was studied in two papers: Setare \cite{seta5} used Kunstatter's approach and used $\omega_R$ to obtain $\bigt A = 24 \pi \hbar \sqrt{ l(l+1) - 1/4}$. Li et.al. \cite{li2} employed the Maggiore's approach and obtained $\bigt A = 8 \pi \hbar$ for the same black hole at the near extreme limit. There are other examples of non-extreme  black holes where the area spectrum was shown to be equally spaced such as, Garfinkle-Horowitz-Strominger black hole\cite{wei2}, and, spinning dilaton black hole in 2 +1 dimensions \cite{fernando6}.



\begin{thebibliography}{99}


\bib{hen1} M. Henneaux, C. Mart$\acute{i}$nez, R. Troncoso, \& J. Zanelli, {\it Asymptotically anti-de Sitter spacetimes and scalar fields with a logarithmic branch}, Phys. Rev. {\bf D70} 044034 (2004)

\bib{hen2} M. Henneaux, C. Mart$\acute{i}$nez, R. Troncoso, \& J. Zanelli, {\it Black holes and asymptotics of 2+1 gravity coupled to a scalar field}, Phys. Rev. {\bf D65}  104007 (2002)

\bib{zan1} C. Mart$\acute{i}$nez, R. Troncoso, \& J. Zanelli, {\it Exact black hole solution with a minimally coupled  scalar field},  Phys. Rev. {\bf D70}  084035 (2004)

\bib{mart2}   C. Mart$\acute{i}$nez, R. Troncoso, {\it Electrically charged black hole with scalar hair}, Phys. Rev. {\bf D74} 064007 (2006)

\bib{xu1}  W. Xu \& L. Zhao, {\it Charged black hole with a scalar hair in (2+1) dimensions},  Phys.  Rev. {\bf D 87}  124008 (2013)

\bib{mart1} C. Mart$\acute{i}$nez, R. Troncoso, \& J. Zanelli, {\it de Sitter black hole with a conformally  coupled  scalar field in four dimensions}, Phys. Rev. {\bf D67}  024008 (2003)

\bib{eli} A. Barlow, D. Doherty, \& E. Winstanley, {\it Thermodynamics of de Sitter black holes with a conformally coupled scalar field}, Phys. Rev. {\bf D 72} 024008 (2005)

\bib{harper} T. J. T. Harper, P. A. Thomas, E. Winstanley, P. M. Young, {\it Instability of four-dimensional de Sitter black hole with a conformally coupled scalar field},  Phys. Rev. {\bf D70} 064023 (2004)


\bib{ana1} A. Anabal$\acute{o}$n \& M. Maeda, {\it New charged black holes with conformal hair}, Phys. Rev. {\bf D81}  041501(2010)

\bib{nada1}  M. Nadalini, L. Vanzo, \& S. Zerbini, {\it Thermodynamical properties of hairy black holes in $n$ spacetime dimensions},  Phys.Rev.{\bf D77} 024047 (2008)


\bib{mart3}  C. Mart$\acute{i}$nez,  \& J. Zanelli, {\it Conformally  dressed black hole in  2+1  dimensions}, Phys. Rev. {\bf D54}  3830 (1996)



\bib{ferra} V. Ferrari \& L. Gualtieri, {\it Quasi-normal modes and gravitational wave astronomy}, Gen. Rel. Grav. {\bf 40} 945 (2008)

\bib{alex} J. Morgan, A. S. Miranda, \& V. T. Zamchin, {\it Electromagnetic quasinormal modes of rotating black strings and the AdS/CFT correspondence}, arXiv: 1302.0536


\bib{cardoso1}  E. Berti, V. Cardoso, \& A, Starinets, {\it Quasinormal modes of black holes and black branes}, Class. Quant. Grav. {\bf 26} 163001 (2009)

\bibitem{kono1}  R. A. Konoplya and A. Zhidenko, {\it Quasinormal modes of black holes: from astrophysics to string theory}, Rev. Mod. Phys. {\bf 83} 793 (2011) 

\bibitem{bek1} J. D. Bekenstein, Lett. Nuovo Cimento {\bf 11}  467 ( 1974)
\bibitem{bek2} J.D. Bekenstein, {\it Quantum black holes as atoms}, arXiv:gr-qc/9710076 



\bibitem{hod} S. Hod, {\it Bohr's correspondence principle and the area spectrum of quantum black holes}, Phys. Rev. Lett. {\bf 81} 4293 (1998)
\bibitem{magi} M. Maggiore, {\it The physical interpretation of the spectrum of black hole quasinormal modes}, Phys. Rev. Lett. {\bf 100} 161301(2008)

\bibitem{elias} E. C. Vagenas, {\it Area spectrum of rotating black holes via the new interpretation of quasinormal modes},  JHEP 0811:073 (2008)
\bib{lopez} A. L$\acute{o}$pez-Ortega, {\it Area spectrum of the D-dimensional de Sitter space-time}, Phys. Lett. {\bf B682} 85 (2009)

\bibitem{fernando1}  S. Fernando, {\it Cold, ultracold and Nariai black holes with quintessence}, Gen. Rel. Grav. {\bf 45} 2053 (2013).


\bib{romans} L. J. Romans, {\it Supersymmetric, cold, lukewarm black holes in cosmological Einstein-Maxwell theory}, Nucl. Phys. {\bf B383} 395 (1992)
\bibitem{paw} J. Matyjasek, P. Sadurski \& D. Tryniecki, {\it Inside the degenerate horizons of the regular black holes},  arXiv:1304.6347

\bibitem{mash} V. Ferrari \& B. Mashhoon, {\it New approach to the quasinormal modes of a black hole},  Phys. Rev. {\bf D30} 295 ( 1984)

\bibitem{kon} R. A. Konoplya,  {\it Quasinormal behavior of the D-dimensional Schwarzschild black hole and higher order WKB approach}, Phys. Rev. {\bf D68}  024018 (2003)

\bibitem{kun} G. Kunstatter, {\it d-Dimensional black hole entropy spectrum from quasi-normal modes}, Phys. Rev. Lett{ \bf 90} 161301 (2003)





\bibitem{ranli} R. Li, {\it Quasinormal modes and entropy spectrum of three dimensional G$\ddot{o}$del black hole}, Int. Jour. Mod. Phys. {\bf D21} 1250014 (2012)
\bib{kwon} Y. Kwon, S. Nam, \& J. Park, {\it Some properties of the de Sitter black holes in three dimensional spacetime}, JHEP {\bf 122} 1311 (2013) 
\bib{bertha} B. Cuadros-Melgar, J. de Oliveira, \& C. E. Pellicer, {\it Stability analysis  and area spectrum of 3-dimensional Lifshitz black holes}, Phys. Rev. {\bf D85} 024014 (2012)
\bib{yun}  Y. S. Myung \& T. Moon, {\it Quasinormal frequencies and thermodynamic quantities for the Lifshitz black holes},  Phys. Rev. {\bf D86} 024006 (2012) 

\bibitem{med} A. J. M. Medved,  {\it On the Kerr quantum area spectrum}, Class. Quan. Grav. {\bf 25} 205014 (2008)



\bib{seta} M. R. Setare, {\it Non-rotating BTZ black hole area spectrum from quasi-normal modes}, Class. Quant. Grav. {\bf 21} 1453 (2004)
\bib{nam1} Y. Kwon \& S. Nam, {\it Area spectrum of the rotating BTZ black hole from quasinormal modes},  Class. Quant. Grav. {\bf 27} 125007 (2010)
\bib{seta3} M. R. Setare, {\it Area spectrum of extremal Reissner-Nordstrom black holes from quasi-normal modes}, Phys. Rev. {\bf D69} 044016 (2004)
\bib{seta4} M. R. Setare \& E. C. Vagenas, {\it Area spectrum of Kerr and extremal Kerr  black holes from quasi-normal modes}, Mod. Phys. Lett {\bf A 20} 1923 (2005)
\bib{seta5} M. R. Setare,  {\it Near extremal Schwarzschild-de Sitter black hole area spectrum  from quasi-normal modes},  Gen. Rel. Grav. {\bf 37} 1411 (2005)
\bibitem{li2} W. Li, L. Xi, \& J. Lu, {\it Area spectrum of near-extremal SdS black holes via the new interpretation of quasinormal modes}, Phys. Lett. {\bf B676} 177 (2009)


\bibitem{wei2} S. Wei, Y. Liu, K. Yang, \& Y. Zhang, {\it Entropy/area spectra of the charged black hole from quasinormal modes}, Phys. Rev. {\bf D81} 104042 (2010)


\bibitem{fernando6} S. Fernando, {\it Spinning dilaton black holes in 2+1 dimensions: quasinormal modes and the area spectrum},  Phys. Rev. {\bf D79} 124026 (2009)








\end{thebibliography}
\end{document}